\documentclass[letterpaper]{article} 
\usepackage{aaai25}  
\usepackage{times}  
\usepackage{helvet}  
\usepackage{courier}  
\usepackage[hyphens]{url}  
\usepackage{graphicx} 
\usepackage{natbib}  
\usepackage{caption} 
\usepackage{algorithm}
\usepackage{algorithmic}

\usepackage{newfloat}
\usepackage{listings}
\DeclareCaptionStyle{ruled}{labelfont=normalfont,labelsep=colon,strut=off} 
\floatstyle{ruled}
\newfloat{listing}{tb}{lst}{}
\floatname{listing}{Listing}
\title{Modality-Independent Graph Neural Networks with Global Transformers for Multimodal Recommendation}
\author{
    Jun Hu\textsuperscript{\rm 1}, Bryan Hooi\textsuperscript{\rm 1}\thanks{Corresponding author.}, Bingsheng He\textsuperscript{\rm 1}, Yinwei Wei\textsuperscript{\rm 2}
}
\affiliations{

    \textsuperscript{\rm 1}School of Computing, National University of Singapore\\
    \textsuperscript{\rm 2}School of Software, Shandong University\\
    jun.hu@nus.edu.sg, \{bhooi, hebs\}@comp.nus.edu.sg, weiyinwei@hotmail.com

}

\usepackage{bibentry}

\usepackage{amsmath}
\usepackage{graphicx}
\usepackage{xcolor}
\usepackage{subfig}
\usepackage{diagbox}
\usepackage{multirow}
\usepackage{algorithm}
\usepackage{listings}

\usepackage{amssymb}

\usepackage{url}

\usepackage{pifont}
\newcommand{\cmark}{\ding{51}}%
\newcommand{\xmark}{\ding{55}}%

\begin{document}

\maketitle

\begin{abstract}

Multimodal recommendation systems can learn users' preferences from existing user-item interactions as well as the semantics of multimodal data associated with items. 
Many existing methods model this through a multimodal user-item graph, approaching multimodal recommendation as a graph learning task. 
Graph Neural Networks (GNNs) have shown promising performance in this domain.
Prior research has capitalized on GNNs' capability to capture neighborhood information within certain receptive fields (typically denoted by the number of hops, $K$) to enrich user and item semantics. 
We observe that the optimal receptive fields for GNNs can vary across different modalities.
In this paper, we propose GNNs with Modality-Independent Receptive Fields, which employ separate GNNs with independent receptive fields for different modalities to enhance performance. 
Our results indicate that the optimal $K$ for certain modalities on specific datasets can be as low as 1 or 2, which may restrict the GNNs' capacity to capture global information.
To address this, we introduce a Sampling-based Global Transformer, which utilizes uniform global sampling to effectively integrate global information for GNNs.
We conduct comprehensive experiments that demonstrate the superiority of our approach over existing methods.
Our code is publicly available at https://github.com/CrawlScript/MIG-GT.

\end{abstract}

%

\section{Introduction}

Recommendation systems predict user preferences by analyzing historical user-item interactions. 
Recently, deep learning has advanced the development of multimodal recommendation systems, which integrate rich multimodal data like texts and images alongside user-item interactions.
Many existing studies~\cite{DBLP:journals/tmm/ZhouM24,DBLP:journals/tkde/SunWZCW24} demonstrate that this utilization enables a richer, more comprehensive understanding of items, thereby enhancing the performance of recommendations. 
Multimodal recommendation systems have been widely used in applications such as e-commerce and micro-video platforms~\cite{DBLP:journals/tois/LiuXGWLH23,DBLP:conf/sigir/ShangGCJWL23,DBLP:conf/mm/CaiQFHX22,DBLP:journals/tois/LiuWLWNC24}.

\begin{figure}[!tp]
\centering\includegraphics[width=2.2in]{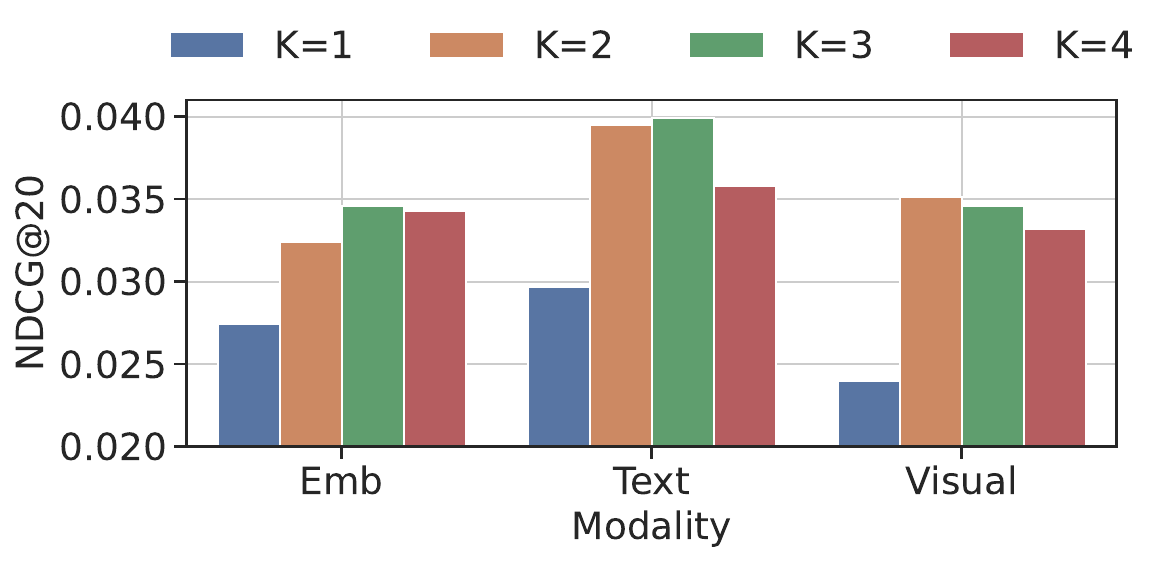}
\vspace{-3mm}
\caption{
Performance of GNNs on Amazon Baby with features of different modalities at varying receptive fields (number of hops, $K$). 
``Emb'' stands for learnable embeddings.
The optimal $K$ is modality-dependent: Emb and Text perform best at $K=3$, while Visual performs best at $K=2$.
}
 \label{fig:hop_performance}
 \vspace{-2mm}
\end{figure}

In recent years, \textbf{GNN-based vertex representation learning} has emerged as a powerful technique in multimedia recommendation systems~\cite{zhou2023comprehensive,DBLP:journals/csur/WuSZXC23,DBLP:journals/tors/GaoZLLQPQCJHL23,DBLP:conf/www/ZhouZLZMWYJ23}.
These approaches use a graph to model the system, typically with graph vertices representing users and items, vertex features encapsulating multimodal data, and graph edges denoting user-item interactions. 
Building on this, recommendation is treated as a task of vertex representation learning. 
By employing GNNs for this purpose, the method effectively utilizes high-order interactions and multimodal data to derive low-dimensional embeddings of users and items. 
These embeddings are then used to compute similarities that reflect user preferences towards specific items, thereby enhancing the performance of the recommendation system.

To handle multimodal data, a typical and effective solution is to apply a separate GNN for each modality, and then pool the representations from these GNNs~\cite{DBLP:conf/mm/WeiWN0HC19}. 
At the feature level, it is obvious that input features of different modalities need to be encoded differently, e.g., using an image encoder for images. 
In this work, we show that the differences between modalities lies not just in how the features should be encoded, but also in \emph{how each modality's information should be propagated over the graph}, i.e. the receptive field used for each modality. 
This issue is overlooked in the recent literature, which does not consider differences at the receptive field level across GNNs for different modalities.

Define the \textbf{receptive field of GNNs} as its number of hops ($K$): existing studies default to setting the same $K$ for GNNs across all modalities.
However, we observe that the optimal receptive field for GNNs differs across modalities.
We conduct experiments by applying GNNs with different $K$ to each modality's features.
When vertices lack features for a modality (e.g., user vertices usually do not have text and visual features), we assign them zero vectors.
Besides text and visual, the learnable embeddings are also treated as a modality.
We conduct the experiment with a state-of-the-art (SOTA) model, MGDN~\cite{10384729} (which generalizes the widely used LightGCN~\cite{DBLP:conf/sigir/0001DWLZ020}).
Results on the Amazon Baby dataset (see Figure~\ref{fig:hop_performance}) show that the learnable embedding and text modalities perform best at $K=3$, while the visual modality performs best at $K=2$, highlighting the benefit of choosing different $K$ for different modalities.

\begin{figure}[!tp]
\centering
\subfloat[
GNNs\label{fig:intro_hop}
]{
\includegraphics[width=1.3in]{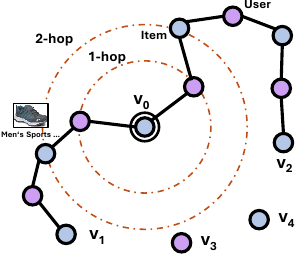}
}
\hspace{4mm}
\subfloat[
Transformers\label{fig:intro_transformer}
]{
\includegraphics[width=1.3in]{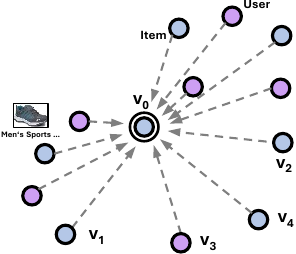}
}
\vspace{-2mm}
\caption{Examples of GNNs and Transformers.}
\label{fig:two_styles} 
\vspace{-5mm}
\end{figure}

Another key consideration is the usage of global information. 
Our experiments (Experiments Section) indicate that the optimal $K$ for certain modalities on specific datasets can be as low as 1 or 2, limiting GNNs' ability to capture global information from vertices across the whole graph.
As shown in Figure~\ref{fig:intro_hop}, when limiting the receptive fields of GNNs to 2 hops, the information of certain vertices, like $v_1$, $v_2$, $v_3$, and $v_4$ is missing for the target vertex $v_0$.
Transformers~\cite{DBLP:conf/nips/VaswaniSPUJGKP17} can potentially capture global information in graphs.
Figure~\ref{fig:intro_transformer} shows that applying a Transformer to all vertices enriches the target vertex $v_0$ with information from all vertices.
However, this is impractical, as Transformers typically require computing attention scores for every vertex pair, resulting in excessive time and space complexity.
To address this, we introduce a Sampling-based Global Transformer, which utilizes uniform global sampling to effectively integrate global information for GNNs.

In this paper, we propose a framework named Modality-Independent Graph Neural Networks with Global Transformers (MIG-GT) for Multimodal Recommendation.
It adopts modality-independent receptive fields to facilitate GNNs on multimodal graphs.
Besides, to exploit the global information, we introduce a Sampling-based Global Transformer, which utilizes uniform global sampling to effectively integrate global information for GNNs.

%
We summarize our key contributions below:
\begin{itemize}
\item We propose GNNs with Modality-Independent Receptive Fields (MIRF), which apply separate GNNs with independent receptive fields (number of hops, $K$) on different modalities, to enhance performance. 
Our results indicate that the optimal $K$ for certain modalities on specific datasets can be as low as 1, which limits the GNN's ability to capture global information.
\item To better capture global information, we introduce a Sampling-based Global Transformer (SGT). 
This module leverages uniform global sampling to effectively incorporate global context into the learning process.
\item We conduct comprehensive experiments that demonstrate the superiority of our method over baselines.
\end{itemize}

\section{Related Work}

\subsection{Graph Neural Networks for Recommendation}

Recent advances in GNNs have facilitated various social media research~\cite{DBLP:conf/www/LiuY0023,DBLP:conf/kdd/ZhuCZGLFDT23,DBLP:journals/tkde/FangZHWX23,DBLP:journals/pami/GaoZX21,DBLP:conf/mm/HanCXZZC21,DBLP:journals/pami/QianXFX23}, with recommendation research being a representative case.
GNN-based methods conceptualize user-item interactions as bipartite graphs, using GNNs to learn embeddings that reflect user preferences. 
GCMC~\cite{DBLP:journals/corr/BergKW17} employs Graph Convolutional Networks (GCNs) to build an autoencoder for recommendation.
PinSage~\cite{DBLP:conf/kdd/YingHCEHL18} uses GNNs with sampling for large datasets.
NGCF~\cite{DBLP:conf/sigir/Wang0WFC19} designs GNNs to enhance interactive signal capture between vertices and neighborhoods.
LightGCN~\cite{DBLP:conf/sigir/0001DWLZ020} optimizes GNNs by forgoing transformations and activation functions within GCN layers, streamlining message passing for effective embedding generation. 
UltraGCN~\cite{DBLP:conf/cikm/MaoZXLWH21} presents a paradigm shift by eschewing explicit GNN operations in favor of a constraint-based loss function. 
ApeGNN~\cite{DBLP:conf/www/ZhangZDWFK023} adaptively aggregates information based on local structures, capturing diverse patterns.
MGDN~\cite{10384729} can generalize LightGCN and offer flexible controls to balance self and neighbor information.

\subsection{Graph Transformers}

Transformers can capture global information but suffer from quadratic complexity w.r.t. vertex count, making them inefficient for large graphs in recommendation tasks.
Recent works, such as SGFormer~\cite{DBLP:conf/nips/WuZYZNJBY23} and Polynormer~\cite{DBLP:conf/iclr/DengYZ24}, address this by removing softmax normalization, reducing complexity to linear.
Both methods combine Graph Transformer outputs with GNN models, differing in their fusion strategies.

\subsection{Multimodal Recommendation}

Initial advancements~\cite{DBLP:conf/aaai/HeM16,DBLP:conf/sigir/LiuWW17} enhanced the Bayesian Personalized Ranking (BPR)~\cite{DBLP:conf/uai/RendleFGS09} by combining learnable embeddings and visual features.
%
%
VECF\cite{DBLP:conf/sigir/ChenCXZ0QZ19} uses VGG~\cite{DBLP:journals/corr/SimonyanZ14a} for image pre-processing and region-specific attention for item visual features.
The advent of GNNs has facilitated multimodal user/item representation learning.
MMGCN~\cite{DBLP:conf/mm/WeiWN0HC19} builds modality-aware graphs and applies separate GNNs to learn modality-specific features, which are then aggregated.
GRCN~\cite{DBLP:conf/mm/WeiWN0C20} progresses this concept by refining user-item graph structures, sieving out misleading connections. 
DualGNN~\cite{DBLP:journals/tmm/WangWYWSN23} introduces a user co-occurrence graph with a feature preference module to capture multimodal item feature dynamics.

Previous methods typically use user-item interaction graphs, capturing item relationships implicitly, while some explicitly model item relationships.
LATTICE~\cite{DBLP:conf/mm/Zhang00WWW21} performs modality-aware structure learning, obtaining item-item structures separately for each modality and then combining them.
FREEDOM~\cite{DBLP:conf/mm/ZhouS23} simplifies the process by freezing the item-item graph structure and denoising the user-item interaction graph.

Most existing approaches focus on denoising or explicitly modeling item-item relations.
Different from them, this paper addresses the task from the perspective of modality-independent receptive fields and global information, showing the necessity of these elements for multimodal recommendation. 
Even without the denoising and explicit modeling of item-item relations, our model already outperforms or matches the performance of SOTA models.

\section{Preliminary}

\subsection{Problem Definition}\label{sec:pre_prob_def}

Let $U$ and $V$ denote the sets of users and items, with sizes $|U|$ and $|V|$, respectively.
We denote the $i$-th user as $u_i$ and the $j$-th item as $v_j$.
Each item in $V$ is associated with multimodal data, including text and an image.
A user-item interaction matrix $B \in \mathbb{R}^{|U| \times |V|}$ represents observed interactions, where $B_{ij}$ is set to 1 or 0 to denote whether the interaction between $u_i$ and $v_j$ has been observed. 
The task is to predict unobserved user preferences over items.
Our model learns a $d$-dimensional vector ($d \ll |U|+|V|$) as the representation for each user/item vertex and uses the dot product between user and item representations to reflect preference scores, with higher scores indicating higher preferences.

\subsection{Multimodal User-Item Graph}

In this paper, we adopt the method of modeling users and items within a single homogeneous graph.
Specifically, the users and items are modeled as vertices of a user-item graph $\mathcal{G}$, consisting of $|N| = |U| + |V|$ vertices, where the first $|U|$ vertices represent users and the subsequent $|V|$ vertices represent items.
The observed user-item interactions are modeled as edges in the graph, thus, the adjacency matrix of the graph $A \in \{0,1\}^{|N| \times |N|}$ is defined as follows:
\vspace{-1mm}
\begin{equation}
\vspace{-1mm}
\small
  A = \begin{pmatrix}
    0                                & B \\
    B' & 0
  \end{pmatrix}
\end{equation}
where $B'$ denotes the transpose of $B$.
For multimodal data, each item vertex has a text feature vector and a visual feature vector extracted via pre-trained models.
Each user vertex is assigned a $d$-dimensional learnable embedding.

\section{Method}

We present our framework, Modality-Independent Graph Neural Networks with Global Transformers (MIG-GT).

\begin{figure*}[!tp]
\centering\includegraphics[width=5.3in]{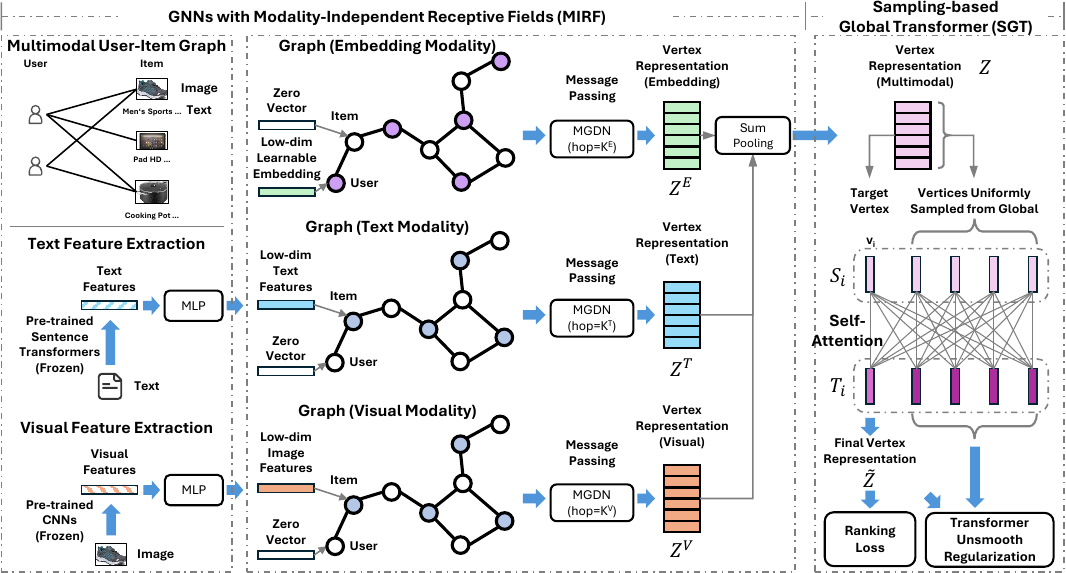}
\vspace{-2mm}
\caption{
Overall Framework of Modality-Independent Graph Neural Networks with Global Transformers (MIG-GT).
}
 \label{fig:overall_framework}
 \vspace{-4mm}
\end{figure*}

\subsection{Overall Framework}

Figure~\ref{fig:overall_framework} provides an overview of our framework.
The upper-left part shows the input graph with users and items as vertices and user-item interactions as edges.
Each item is associated multimodal data, including text and an image.
Note that our method is applied directly to the original user-item interaction graph without constructing new graphs.

Our proposed framework mainly consists of two components:
(1) \textbf{Modality-Independent Receptive Fields (MIRF)} applies separate GNNs with independent receptive fields for data of different modalities in the graph.
For each GNN, first an MLP is used to encode vertex features of the corresponding modality into $d$-dimensional feature vectors, which is a common operation in existing works.~\cite{DBLP:conf/mm/WeiWN0HC19}
Then we perform message propagation with the encoded features, and different from existing work, we propose to use modality-independent receptive fields for different modalities.
The independent receptive fields for different modalities, learnable embedding, text, and visual, denoted as $K^{(E)}$, $K^{(T)}$, and $K^{(V)}$ are selected based on validation set performance.
(2) \textbf{Sampling-based Global Transformer (SGT)} is designed to exploit global information.
SGT performs self-attention using a few vertices uniformly sampled from the entire graph to enrich vertex representations. 
Unlike typical Transformers, which compute attention scores for every pair of vertices—resulting in significant time and space complexity—SGT only needs to compute the attention scores between each vertex and the few sampled vertices, thereby enhancing efficiency.
We also propose a Transformer Unsmooth Regularization (TUR) for optimization.

\subsection{Modality-Independent Receptive Fields}

We apply separate GNNs to features from different modalities.
Each GNN follows an encode-then-propagate framework ~\cite{DBLP:conf/iclr/KlicperaBG19}, using an MLP to encode vertex features into $d$-dimensional vectors for message propagation on the graph.
($d \ll |N|$ is the dimensionality of the final vertex representations.)
For message propagation, we utilize MGDN, which propagates vertex features without feature transformations.

Specifically, we use $X^{(M)} \in \mathbb{R}^{|N| \times d^{(M)}}$ to denote the raw vertex feature matrix of modality $M$, where $|N|$ is the number of vertices, and $d^{(M)}$ is the feature dimension specific to modality $M$.
$X^{(E)}$, $X^{(T)}$, and $X^{(V)}$ represent the learnable embedding, text, and visual modalities, respectively. 
The encoded feature vectors by the MLP are denoted as:
\begin{equation}
\small
    \tilde{X}^{(M)} = \mathrm{MLP}(X^{(M)}) \in \mathbb{R}^{|N| \times d}
\end{equation}
One exception is the learnable embedding modality, where $X^{(E)} \in \mathbb{R}^{|N| \times d}$ can be directly optimized. 
Therefore, it does not require an extra MLP, and $\tilde{X}^{(E)} = X^{(E)}$.
Note that treating each modality independently may cause certain vertices to have missing features.
For example, when dealing with either text or visual modalities, the user vertices may lack features under certain modalities.
In such cases, we simply assign zero vectors as the encoded feature vectors for these featureless vertices (in $\tilde{X}^{(M)}$), with their dimensionality matching that of other vertices' features.

For message propagation, we use MGDN separately for each modality with different receptive fields $K^{(M)}$ for each.
We first compute a normalized adjacency matrix for MGDN, shared across all modalities, $\hat{A} = \tilde{D}^{-\frac{1}{2}}\tilde{A}\tilde{D}^{-\frac{1}{2}}$, where $\tilde{A} = A + I$ and  $\tilde{D}_{ii} = \sum_{j=0}^{N} \tilde{A}_{ij}$ is the degree matrix of $\tilde{A}$.
With it, MGDN learns vertex representations by incorporating neighbor information within $K^{(M)}$ hops as follows: 
\vspace{-2mm}
\begin{equation} \label{eq:mgdn_coefs}
\vspace{-2mm}
\small
\begin{split}
  Z^{(M)} 
  & = f_{MGDN}(\tilde{X}^{(M)}, A) \\
  & = (\beta^K \hat{A}^{K^{(M)}} + \sum_{k=0}^{K^{(M)}-1} \alpha \beta^{k} \hat{A}^{k}) \tilde{X}^{(M)} / \Gamma
\end{split}
\end{equation}
where $\Gamma$ is used for normalization to ensure the sum of coefficients of $\hat{A}^{k} \tilde{X}^{(M)}$ in Equation~\ref{eq:mgdn_coefs} is 1.0:
\vspace{-2mm}
\begin{equation}
\vspace{-2mm}
\small
  \Gamma = \beta^{K^{(M)}} + \sum_{k=0}^{K^{(M)}-1} \alpha \beta^{k}
\end{equation}
For efficiency, $Z^{(M)}$ is computed in a step-wise manner:
\vspace{-1mm}
\begin{equation}\label{eq:iter_init}
\vspace{-1mm}
\small
  H^{(M,0)} = \tilde{X}^{(M)}
\end{equation}
\begin{equation}\label{eq:iter_recur}
\vspace{-1mm}
\small
  H^{(M,k)} = \beta \hat{A} H^{(M,k-1)} + \alpha H^{(M,0)}
\end{equation}
\begin{equation}\label{eq:iter_final}
\vspace{-1mm}
\small
  Z^{(M)} = H^{(M,K^{(M)})} / \Gamma
\end{equation}
In each iteration, the updated vertex representations are formed by incorporating both the propagation result, $\hat{A} H^{(M,k-1)}$, and the input embeddings, $H^{(M,0)} = \tilde{X}^{(M)}$.
The relative importance of these components is modulated by the hyperparameters $\beta$ and $\alpha$.

After obtaining modality-independent vertex representations $Z^{(E)}$, $Z^{(T)}$, and $Z^{(V)}$, we perform sum-pooling to obtain the multimodal vertex representations $Z \in \mathbb{R}^{|N| \times d}$:
\vspace{-1mm}
\begin{equation}
\vspace{-1mm}
\small
Z = Z^{(E)} + Z^{(T)} + Z^{(V)}
\end{equation}
We define $z_i$ to denote the $i$-th row in $Z$, representing the multimodal representation of the $i$-th vertex.

For selecting modality-independent receptive fields $K^{(M)}$ ($K^{(E)}$, $K^{(T)}$, and $K^{(V)}$), we show in the experiments section that grid search on validation datasets is feasible, with the impact of these hyperparameters on validation datasets being almost consistent with that on test datasets.

\subsection{Sampling-Based Global Transformer}\label{sec:method_gt}

Typical Transformers require computing the attention scores between every pair of samples, thus making it impractical to apply them directly to graphs.
To alleviate this, for each vertex $z_i$, we uniformly sample $C$ vertex representations from $Z$, and apply Transformers only on the current vertex and the sampled vertices.
Note that, for each vertex, $C$ vertices are sampled independently at each training step.
Therefore, throughout the training, each vertex is likely to be sampled along with many vertices uniformly sampled from the global set, capturing more global information.
This global information will then be learned by the MLP encoder of the GNNs and the learnable embeddings.

Formally, given $z_i$, we construct a matrix $S_i \in \mathbb{R}^{(C+1) \times d}$, where the first row $s_{i1} = z_i$ represents the representation of the $i$-th vertex. 
Each subsequent row $s_{ij}$ (for $1 < j \leq C + 1$) is assigned a vertex representation $z_k$ from $Z$ denoted as $s_{ij} = z_k$, and $k \sim \text{Uniform}(1, |N|)$.

We apply a simplified Transformer on $S_i$ to perform self-attention to enrich the semantics of the vertex representations in $S_i$, obtaining $T_i \in \mathbb{R}^{(C+1) \times d}$:
\vspace{-2mm}
\begin{equation}
\vspace{-2mm}
\small
    T_i = (1 - \gamma) \mathrm{softmax}\left(\frac{\mathcal{Q}\mathcal{K}^\top}{\sqrt{d}}\right)\mathcal{V} + \gamma S_i
\end{equation}
where $\mathrm{softmax}$ denotes the row-wise softmax normalization, and $0 \leq \gamma \leq 1$ is a hyperparameter that modifies Transformer's residual connection to offer more flexibility, allowing for adjustable integration of the Transformer's output with the original input.
$\mathcal{Q} = S_i W^{(\mathcal{Q})}$, $\mathcal{K} = S_iW^{(\mathcal{K})}$, and $\mathcal{V} = S_i$, where $W^{(\mathcal{Q})} \in \mathbb{R}^{d \times d^{(\text{att})}}$ and $W^{(\mathcal{K})} \in \mathbb{R}^{d \times d^{(\text{att})}}$ are learnable parameters, and $d^{(\text{att})}$ is a hyperparameter.

After self-attention, we extract the first row of $T_i$ as the final representation of the $i$-th vertex.
We denote the final vertex representation matrix as $\tilde{Z} \in \mathbb{R}^{|N| \times d}$, where $\tilde{z}_i$ represents the $i$-th vertex, thus $\tilde{z}_i = T_{i1}$.
Note that the remaining rows $\{T_{ij} | j > 1\}$ are not taken as the final vertex representations but are used for regularization, introduced later.
Besides, we use $\tilde{u}_i=\tilde{z}_i$ and $\tilde{v}_j=\tilde{z}_{j+|U|}$ to denote the final vertex representation of the $i$-th user and $j$-th item.

\textbf{Transformer Unsmooth Regularization}
Our method samples vertices globally and uses the Transformer to enrich them by extracting complementary information from each other. 
This process may cause an issue known as smoothing, making it difficult to distinguish the representations of certain vertices. 
Given the $i$-th vertex and the corresponding output of the Transformer, $T_i$, it contains the enriched representation of not only the $i$-th vertex but also $C$ uniformly sampled vertices. 
Since they complement each other via self-attention, their representations are likely to become smooth and indistinguishable from one another. 
Our unsmooth regularization aims to ensure our model can still distinguish between them. 
This regularization relies on edges in graphs. 
Sampling one of the $i$-th vertex's neighbors, denoted as the $k$-th vertex (that is, $A_{ik}=1$), we take its final representation $\tilde{z}_k$ and apply the regularization as follows:
\begin{equation}
\small
    \mathcal{L}_{TUR} = - \sum_{A_{ik}=1}  \log \left( \frac{\exp(\tilde{z}_k' T_{i1})}{\sum_{j=1}^{C+1} \exp(\tilde{z}_k' T_{ij})} \right)
\end{equation}

Our experiments show that with a small number of vertices sampled from the global graph ($C \leq 20$), our model can already achieve significant performance improvements. 
Note that, unlike a typical Transformer, which requires computing attention scores between each vertex and all vertices in the graph—resulting in significant time and space complexity—our method computes attention scores only between each vertex and the $C$ sampled vertices, where $C$ is very small, thereby enhancing efficiency.

\subsection{Model Optimization}

We optimize a combined loss with Adam optimizer~\cite{DBLP:journals/corr/KingmaB14}, where the combined loss is as follows:
\begin{equation}
\small
\mathcal{L}_{rec} = \mathcal{L}_{rank}(\tilde{Z})+ \mathcal{L}_{TUR}(\tilde{Z}, T) + \Psi_{L2}\mathcal{L}_{L2}(\tilde{Z})
\end{equation}
where $\mathcal{L}_{rank}(\tilde{Z})$ is the ranking loss, and $\mathcal{L}_{L2}(\tilde{Z})$ is the L2 regularization with coefficient $\Psi_{L2}$.
For the ranking loss, we adopt the popular BPR loss~\cite{DBLP:conf/uai/RendleFGS09}:
\begin{equation}\label{eq:bpr}
\small
  \mathcal{L}_{BPR} = - \sum_{B_{ij}=1} \mathbb{E}_{v_k \sim p(v)} \log \sigma (\tilde{u}_i' \tilde{v}_j - \tilde{u}_i' \tilde{v}_k)
\end{equation}
where $\sigma$ denotes the sigmoid activation function, and $v_k \sim p(v)$ represents a vertex randomly sampled from the graph.

\section{Experiments}

We conduct experiments on three public datasets using a Linux system with two Intel(R) Xeon(R) CPU E5-2690 v4 @ 2.60GHz processors, 128GB of RAM, and a GeForce GTX 1080 Ti GPU (11GB).
The model is implemented via PyTorch~\footnote{\url{https://pytorch.org/}} and DGL~\cite{wang2020deep}, with our code included in the supplementary materials.

\subsection{Datasets}

Following prior work~\cite{DBLP:conf/mm/Zhang00WWW21,DBLP:conf/mm/ZhouS23}, we conduct our studies on three public datasets from the Amazon review datasets~\cite{DBLP:conf/www/HeM16}, specifically abbreviated as Baby, Sports, and Clothing, respectively.
These datasets provide multimodal data (textual and visual) for the items and vary in item count per category.
We utilize the preprocessed data from previous studies, where the raw data for each category was filtered using a 5-core threshold for both products and users.
Regarding multimodal features, we adopt the text and visual embeddings extracted and published by prior work~\cite{DBLP:conf/mm/ZhouS23}, with visual features as 4,096-dimensional embeddings obtained from pre-trained Convolutional Neural Networks, and text features as 384-dimensional embeddings from sentence-transformers, derived from item titles, descriptions, categories, and brands.
Dataset statistics are shown in Table~\ref{tab:dataset_statistics}.

\begin{table}[!tp]
\centering

\scalebox{0.85}{
\begin{tabular}{|l|c|c|c|c|}
\hline
Dataset    & Users    & Items    & Interactions & Sparsity \\ \hline
Baby       & 19,445   & 7,050    & 160,792         & 99.88\%  \\
Sports     & 35,598   & 18,357   & 296,337         & 99.95\%  \\
Clothing   & 39,387   & 23,033   & 278,677         & 99.97\%  \\ \hline
\end{tabular}
}
\vspace{-2mm}
\caption{Statistics of the three datasets.}
\label{tab:dataset_statistics}
\vspace{-4mm}
\end{table}

\begin{table*}[!tp]
\centering
\scalebox{0.72}{
\begin{tabular}{l|cc|cccc|cccc|cccc}
\hline
 & & &\multicolumn{4}{c|}{Baby} & \multicolumn{4}{c|}{Sports} & \multicolumn{4}{c}{Clothing} \\
\hline
Method   & Multimodal 
                  & GNN    & R@10 & R@20 & N@10 & N@20 & R@10 & R@20 & N@10 & N@20 & R@10 & R@20 & N@10 & N@20 \\
\hline
MF       & \xmark & \xmark & 0.0357 & 0.0575 & 0.0192 & 0.0249 & 0.0432 & 0.0653 & 0.0241 & 0.0298 & 0.0206 & 0.0303 & 0.0114 & 0.0138 \\
LightGCN & \xmark & \cmark & 0.0479 & 0.0754 & 0.0257 & 0.0328 & 0.0569 & 0.0864 & 0.0311 & 0.0387 & 0.0361 & 0.0544 & 0.0197 & 0.0243 \\
ApeGNN	 & \xmark & \cmark & 0.0501	& 0.0775 & 0.0267 & 0.0338 & 0.0608 & 0.0892 & 0.0333 & 0.0407 & 0.0378 & 0.0538 & 0.0204 & 0.0244 \\
MGDN     & \xmark & \cmark & 0.0495 & 0.0783 & 0.0272 & 0.0346 & 0.0614 & 0.0932 & 0.0340 & 0.0422 & 0.0362 & 0.0551 & 0.0199 & 0.0247\\
\hline
VBPR     & \cmark & \xmark & 0.0423 & 0.0663 & 0.0223 & 0.0284 & 0.0558 & 0.0856 & 0.0307 & 0.0384 & 0.0281 & 0.0415 & 0.0158 & 0.0192 \\
MMGCN    & \cmark & \cmark & 0.0421 & 0.0660 & 0.0220 & 0.0282 & 0.0401 & 0.0636 & 0.0209 & 0.0270 & 0.0227 & 0.0361 & 0.0154 & 0.0154 \\
GRCN     & \cmark & \cmark & 0.0532 & 0.0824 & 0.0282 & 0.0358 & 0.0599 & 0.0919 & 0.0330 & 0.0413 & 0.0421 & 0.0657 & 0.0224 & 0.0284 \\
DualGNN  & \cmark & \cmark & 0.0513 & 0.0803 & 0.0278 & 0.0352 & 0.0588 & 0.0899 & 0.0324 & 0.0404 & 0.0452 & 0.0675 & 0.0242 & 0.0298 \\
SLMRec   & \cmark & \cmark & 0.0521 & 0.0772 & 0.0289 & 0.0354 & 0.0663 & 0.0990 & 0.0365 & 0.0450 & 0.0442 & 0.0659 & 0.0241 & 0.0296 \\
LATTICE  & \cmark & \cmark & 0.0547 & 0.0850 & 0.0292 & 0.0370 & 0.0620 & 0.0953 & 0.0335 & 0.0421 & 0.0492 & 0.0733 & 0.0268 & 0.0330 \\
FREEDOM  & \cmark & \cmark & 0.0627 & 0.0992 & 0.0330 & 0.0424 & 0.0717 & 0.1089 & 0.0385 & 0.0481 & 0.0626 & 0.0932 & 0.0338 & 0.0416 \\
\hline
MIG-GT  & \cmark & \cmark & \textbf{0.0665} 
                                   & \textbf{0.1021} 
                                            & \textbf{0.0361} 
                                                    & \textbf{0.0452} 
                                                               & \textbf{0.0753}
                                                                        & \textbf{0.1130} 
                                                                                & \textbf{ 0.0414} 
                                                                                        & \textbf{0.0511} 
                                                                                                  & \textbf{0.0636} 
                                                                                                            & \textbf{0.0934}
                                                                                                                    & \textbf{0.0347} 
                                                                                                                            & \textbf{0.0422}
\\\hline
Improv.  &        &        & 6.06\%  & 2.92\%& 9.39\% & 6.6\%  & 5.02\%  & 3.76\% & 7.53\% & 6.24\% &  1.6\%  & 0.21\% & 2.66\% & 1.44\%     \\ 

\hline
\end{tabular}
}
\vspace{-2mm}
\caption{Performance comparison of different recommendation models.}
\label{tab:exp_overall_performance}
\vspace{-5mm}
\end{table*}

\subsection{Baselines}

%
All baselines utilize BPR as the ranking loss.
The first set of baselines comprises models that rely solely on user-item interactions and do not incorporate multimodal data: \textbf{MF}~\cite{DBLP:journals/computer/KorenBV09}, \textbf{LightGCN}~\cite{DBLP:conf/sigir/0001DWLZ020}, \textbf{ApeGNN}~\cite{DBLP:conf/www/ZhangZDWFK023}, and \textbf{MGDN}~\cite{10384729}.
The second set includes multimodal recommendation models that leverage both user-item interactions and multimodal data: \textbf{VBPR}~\cite{DBLP:conf/aaai/HeM16}, \textbf{MMGCN}~\cite{DBLP:conf/mm/WeiWN0HC19}, \textbf{GRCN}~\cite{DBLP:conf/mm/WeiWN0C20}, \textbf{DualGNN}~\cite{DBLP:journals/tmm/WangWYWSN23}, \textbf{SLMRec}~\cite{DBLP:journals/tmm/TaoLXWYHC23}, \textbf{LATTICE}~\cite{DBLP:conf/mm/Zhang00WWW21}, and \textbf{FREEDOM}~\cite{DBLP:conf/mm/ZhouS23}.

\subsection{Model Evaluation and Parameter Settings}

To ensure a fair comparison, we adopt the evaluation settings from previous studies~\cite{DBLP:journals/tmm/TaoLXWYHC23,DBLP:journals/tmm/WangWYWSN23,DBLP:conf/mm/Zhang00WWW21,DBLP:conf/mm/ZhouS23}. 
Our evaluation criteria include two widely-used metrics, abbreviated as R for recall and N for Normalized Discounted Cumulative Gain (NDCG). 
We report these metrics for top 10 and top 20 recommendations, denoted as R@10, R@20, N@10, and N@20.
Regarding data split, we allocated 80\% of known user interactions for training, 10\% for validation, and the remaining 10\% for testing.
The reported performance is the mean results obtained using five different random seeds.

In terms of hyperparameters, we tune each hyperparameter within  a range, and take the combination achieving best performance in the validation set, reporting the corresponding test performance.
For the independent receptive fields, we search number $K^{(M)} \leq 4$.
For the $\gamma$ in the transformer, we search within 0.8 and 0.9.
The learning rate and L2 regularization coefficient are searched from $\{1 \times 10^{-2}, 1 \times 10^{-3}\}$ and $\{1 \times 10^{-4}, 1 \times 10^{-5}\}$, respectively.

\subsection{Performance Analysis}

We report the performance of baseline methods and our method in Table~\ref{tab:exp_overall_performance}. 
Some baseline results are directly cited from the original literature, while others, which may slightly differ, are derived from running the provided source code without modifications.
The column `Multimodal' indicates whether the method utilizes multimodal data, and the column `GNN' denotes whether it is GNN-based.
From the reported performance, we make the following observations:
\begin{itemize}
\item The methods exploiting multimodal data generally demonstrate an advantage over those that do not utilize multimodal data. Notably, VBPR, which extends MF via multimodal features, shows a significant improvement in performance, underscoring the importance of leveraging multimodal data in recommendation systems.
\item 
GNN-based methods consistently outperform non-GNN methods (MF and VBPR), with or without multimodal data, highlighting the efficacy of GNNs in this domain.
\item MMGCN is an early and typical study on applying GNNs for multimodal recommendation, proposing to apply different GNNs to different modalities separately.
Compared to MMGCN, GRCN, DualGNN, and SLMRec consider more nuanced properties, like noisy user-item interactions, resulting in better performance. 
This shows that there is room to improve GNNs based on the specific properties in multimodal recommendation systems.
\item Besides explicit user-item interactions, LATTICE and FREEDOM learn implicit item-item relationships and build graphs to explicitly utilize them. 
FREEDOM also introduces a denoising mechanism for the learned item-item relations.
Both models outperform most other baselines, demonstrating that explicitly modeling item-item relationships is an effective alternative approach.
\item MIG-GT outperforms the SOTA baseline (FREEDOM) with an improvement of around 5\% on two datasets (Baby and Sports) and slightly outperforms FREEDOM on Clothing.
Note that, unlike other GNN baselines, MIG-GT does not rely on commonly used components like denoising and explicit modeling of item-item relations. 
It relies solely on our modality-independent receptive fields and a sampling-based Global Transformer, yet it still outperforms or matches the performance of baselines, showing the effectiveness of our method.

\end{itemize}

\subsection{Detailed Analysis}

\subsubsection{Impact and Selection of Modality-Independent Receptive Fields}\label{sec:impact_mirf}

Firstly, we examine the impact of modality-independent receptive fields, denoted as ${K^{(M)}}$, which include ${K^{(E)}}$, ${K^{(T)}}$, and ${K^{(V)}}$. 
Given the extensive number of possible combinations, we present a subset of these on the Amazon Baby dataset for brevity. 
In each figure, we fix ${K^{(M)}}$ of one modality to the optimal value  
and vary the others from 1 to 4, resulting in 16 combinations. We report both validation and test performance for each combination.

\begin{figure}[!tp]
\centering
\subfloat[
Performance on Valid Set.\label{fig:impact_mirf_valid_kt_kv}
]{
\includegraphics[width=1.6in]{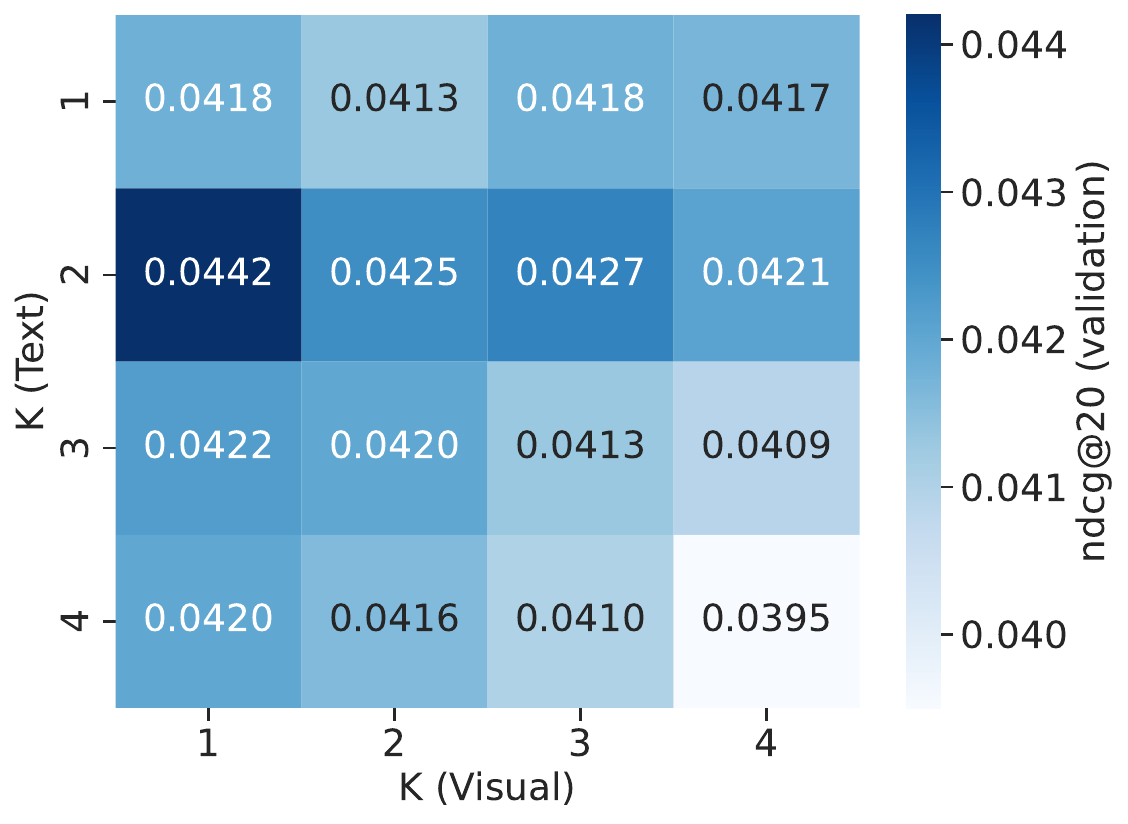}
}
\subfloat[
Performance on Test Set.\label{fig:impact_mirf_test_kt_kv}
]{
\includegraphics[width=1.6in]{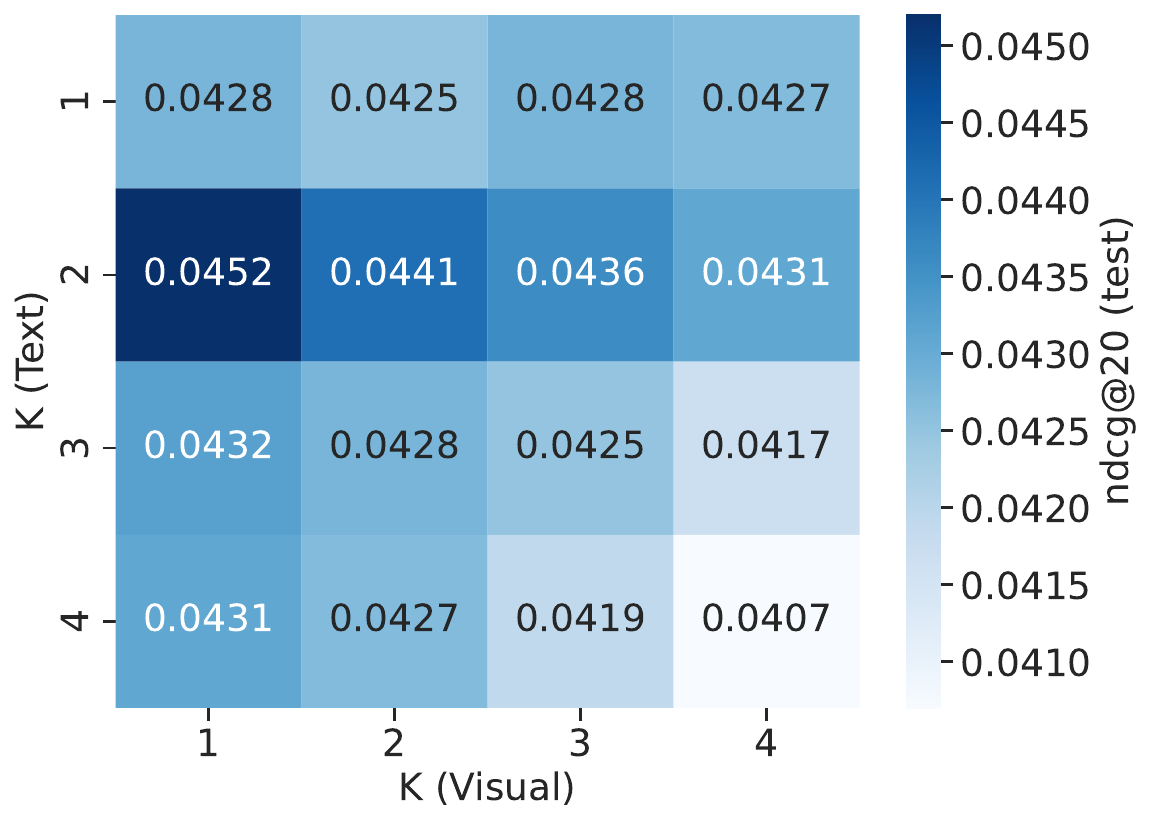}
}
\vspace{-3mm}
\caption{Heatmaps showing the NDCG@20 scores for different combinations of ${K^{(T)}}$ and ${K^{(V)}}$.}
\label{fig:impact_mirf_kt_kv} 
\vspace{-4mm}
\end{figure}

\begin{figure}[!tp]
\centering
\subfloat[
Performance on Valid Set.\label{fig:impact_mirf_valid_ke_kv}
]{
\includegraphics[width=1.6in]{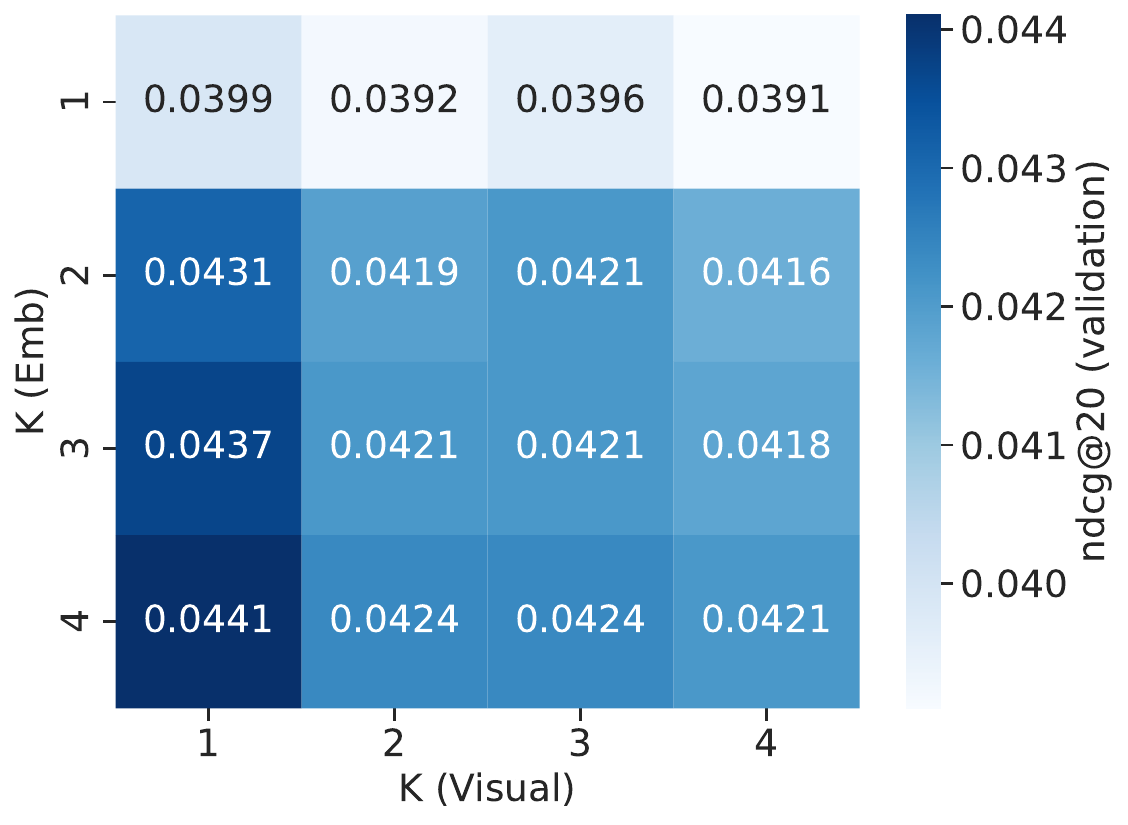}
}
\subfloat[
Performance on Test Set.\label{fig:impact_mirf_test_ke_kv}
]{
\includegraphics[width=1.6in]{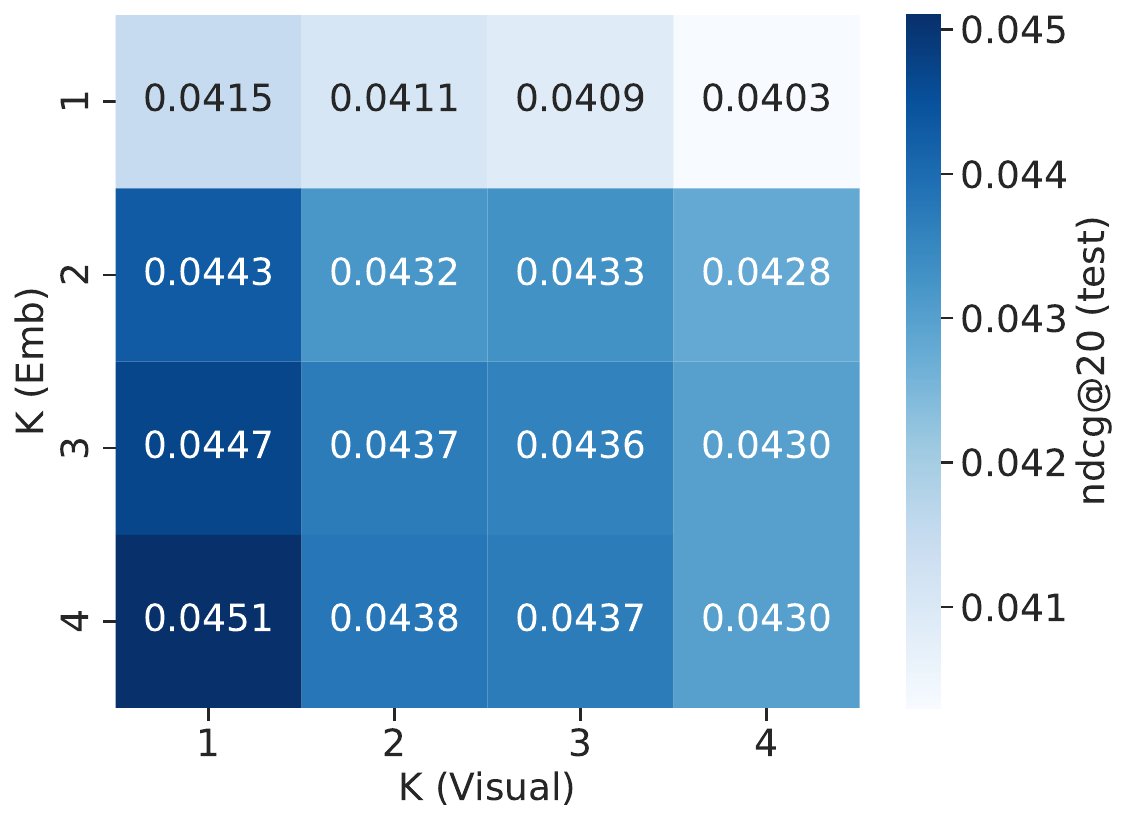}
}
\vspace{-3mm}
\caption{Heatmaps showing the NDCG@20 scores for different combinations of ${K^{(E)}}$ and ${K^{(V)}}$.}
\label{fig:impact_mirf_ke_kv} 
\vspace{-5mm}
\end{figure}

To clearly visualize the impact of MIRFs, we use several heatmaps in Figure~\ref{fig:impact_mirf_kt_kv} and Figure~\ref{fig:impact_mirf_ke_kv}, with the x-axis and y-axis representing the numbers of two modality receptive fields, respectively. 
Each cell in the heatmap displays the performance (ndcg@20), highlighting performance variations across different combinations. 
In Figure~\ref{fig:impact_mirf_kt_kv}, we fix ${K^{(E)}}=4$ and vary ${K^{(T)}}$ and ${K^{(V)}}$. 
Another set of heatmaps in Figure~\ref{fig:impact_mirf_ke_kv} maintains ${K^{(T)}}=2$ while varying ${K^{(E)}}$ and ${K^{(V)}}$. 
The results indicate that different combinations yield different performances, with the optimal MIRF configuration (which diverges from the identity receptive field setting across all modalities) achieving the best performance.

\textbf{Selection of MIRFs.} 
Furthermore, we demonstrate the feasibility of using grid search on the validation dataset for selecting MIRFs. 
Figure~\ref{fig:impact_mirf_valid_kt_kv} and Figure~\ref{fig:impact_mirf_valid_ke_kv} depict the heatmaps generated from the validation dataset, while Figure~\ref{fig:impact_mirf_test_kt_kv} and Figure~\ref{fig:impact_mirf_test_ke_kv} correspond to the test dataset. 
Comparing the two sets of heatmaps allows us to assess the consistency of the hyperparameters' impact between the validation and test datasets. 
Our findings confirm that the patterns observed during validation are generally representative of the test phase, validating grid search as a viable method for selecting independent receptive fields for different modalities.

\subsubsection{Impact of Sampling-Based Global Transformers}\label{sec:impact_gt}
\begin{figure}[!tp]
\centering
\subfloat[
recall@10
]{
\includegraphics[width=1.5in]{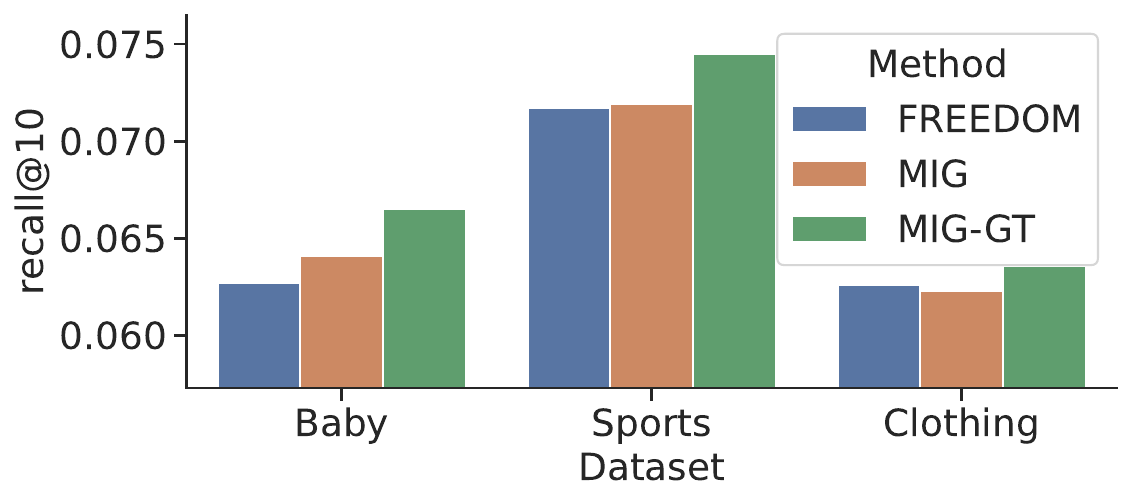}
}
\subfloat[
recall@20
]{
\includegraphics[width=1.5in]{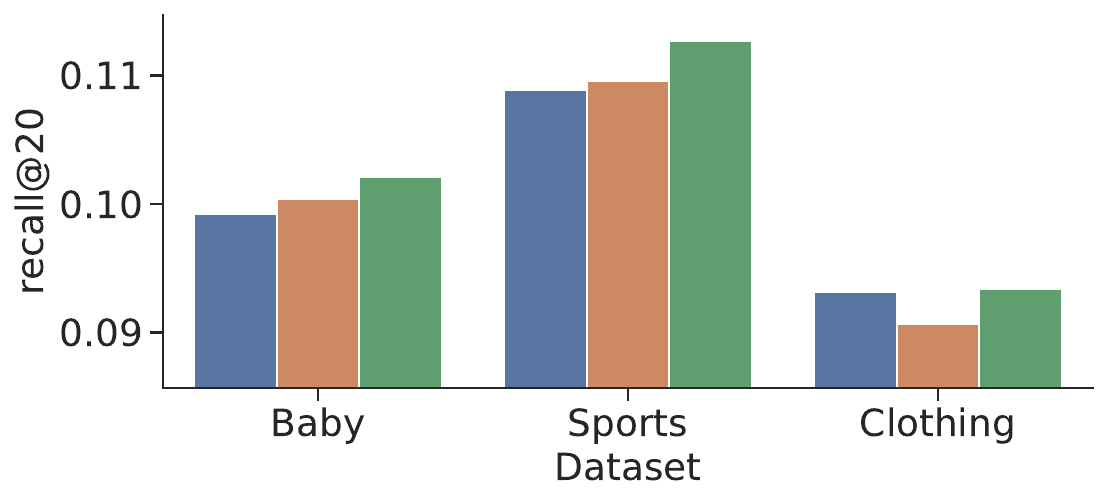}
}
\\
\subfloat[
ndcg@10
]{
\includegraphics[width=1.5in]{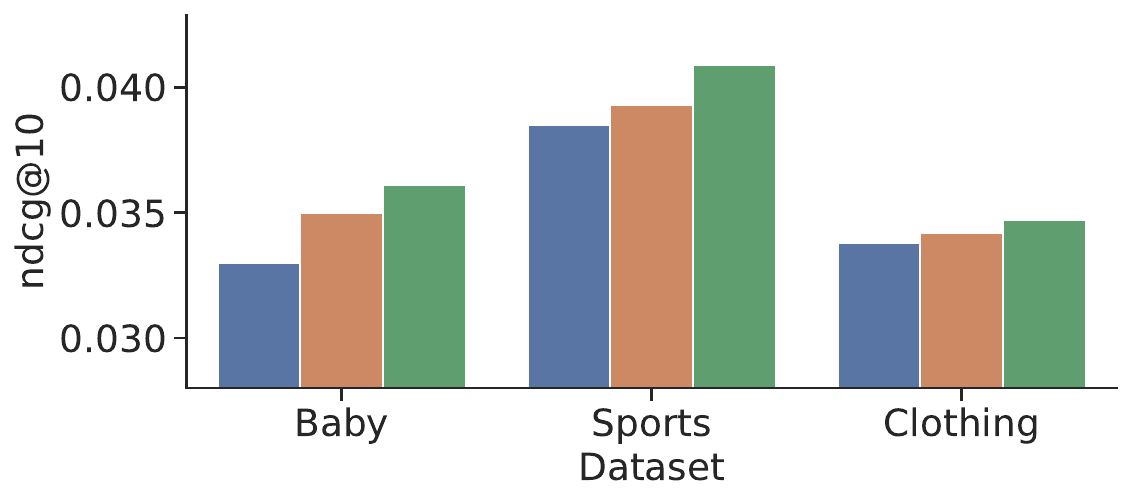}
}
\subfloat[
ndcg@20
]{
\includegraphics[width=1.5in]{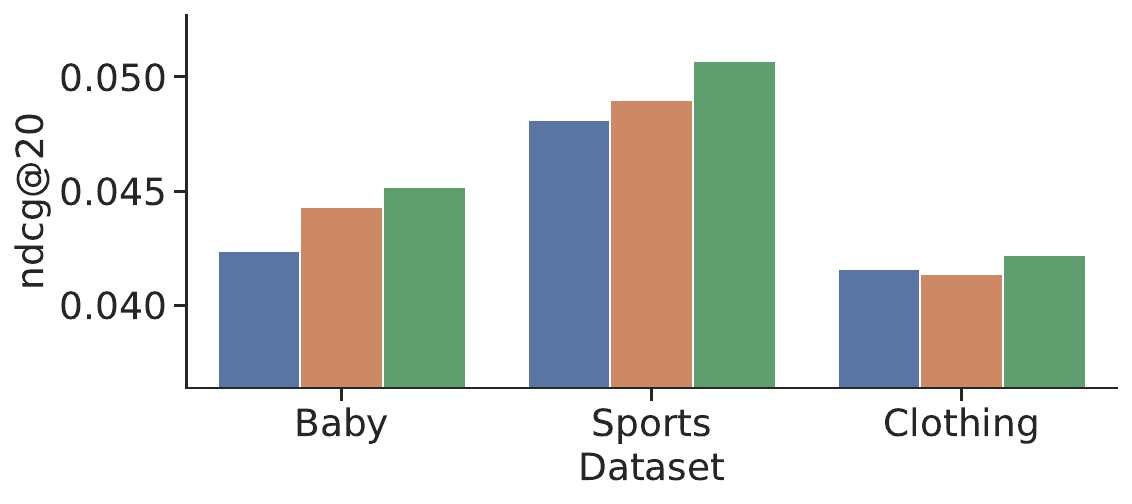}
}
\vspace{-2mm}
\caption{Impact of Sampling-based Global Transformers.}
\label{fig:ablation_gtg} 
\vspace{-2mm}
\end{figure}

We perform ablation tests with a variant of our model, MIG, which removes SGT and retains only the MIRF components.
We compare our model against the variant in Figure~\ref{fig:ablation_gtg}, including the performance of the SOTA method FREEDOM for better comparison, to show the effectiveness of SGT.
The results show that MIG, even without SGT, already outperforms FREEDOM on Baby and Sports.
With SGT, MIG-GT further enhances MIG's performance.
On Clothing, although MIG is slightly outperformed by FREEDOM, with SGT, MIG-GT enhances it to achieve better performance than FREEDOM, showing the effectiveness of SGT.

Additionally, we replace SGT with existing Graph Transformer methods, SGFormer and Polynormer, to construct variants MIG-SGFormer and MIG-Polynormer, and compare them with MIG-GT in Table~\ref{tab:exp_impact_diff_gts}. 
The results show that MIG-GT outperforms these variants, demonstrating the effectiveness of our sampling-based approach for developing Global Transformers in recommendation contexts.

\begin{table}[!tp]
\centering
\scalebox{0.72}{
\begin{tabular}{l|cc|cc|cc}
\hline
 & \multicolumn{2}{c|}{Baby} & \multicolumn{2}{c|}{Sports} & \multicolumn{2}{c}{Clothing} \\
\hline
Method          &  R@20  & N@20   &  R@20 & N@20  &  R@20 & N@20  \\
\hline

MIG-SGFormer	& 0.0863 & 0.0376 &	0.0887 & 0.0392 & 0.0827 & 0.0363 \\
MIG-Polynormer	& 0.0997 & 0.0436 &	0.1048 & 0.0461 & 0.0864 & 0.0386 \\ \hline
MIG-GT	        & \textbf{0.1021} & \textbf{0.0452} &	\textbf{0.1130} & \textbf{0.0511} & \textbf{0.0934} & \textbf{0.0422} \\

\hline

\hline
\end{tabular}
}
\vspace{-2mm}
\caption{Impact of Different Global Transformers.}
\label{tab:exp_impact_diff_gts}
\vspace{-2mm}
\end{table}

\begin{figure}[!tp]
\centering
\subfloat[
recall@20
]{
\includegraphics[width=1.5in]{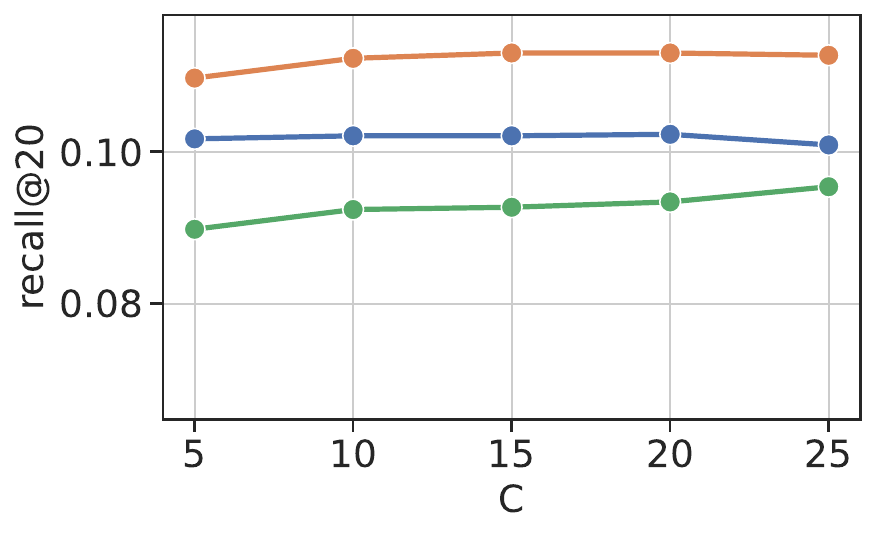}
}
\subfloat[
ndcg@20
]{
\includegraphics[width=1.5in]{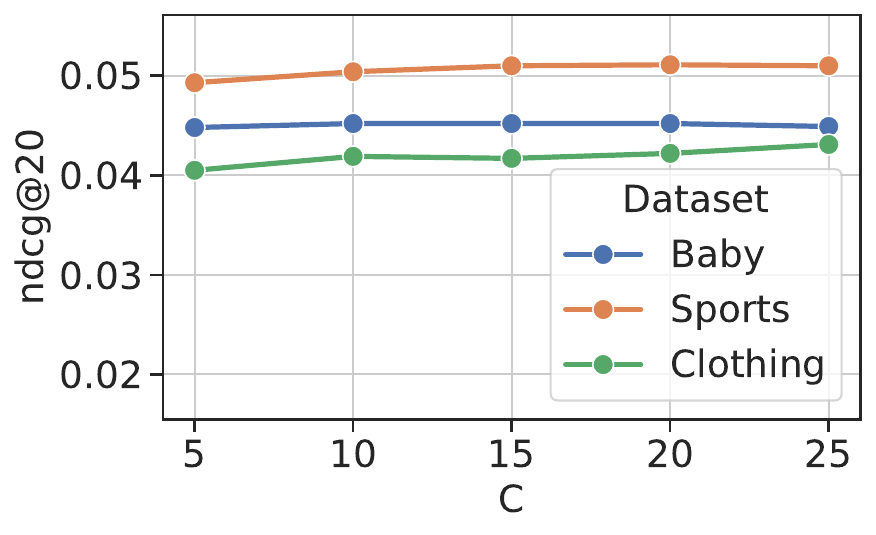}
}
\vspace{-2mm}
\caption{Impact of Number of Global Samples ($C$) for SGT.}
\label{fig:impact_num_samples} 
\vspace{-2mm}
\end{figure}

\subsubsection{Impact of Number of Global Samples for SGT}\label{sec:impact_num_samples}

To investigate the impact of the number of global samples $C$ for SGT, we vary it from 5 to 25 and report the corresponding performance. 
Results in Figure~\ref{fig:impact_num_samples} show that when increasing from 5 to 10, we observe a performance improvement across datasets. Further increasing it beyond 10, performance increments can be observed on certain datasets. 
Overall, this demonstrates that with only 10 or 20 global samples, our SGT can significantly improve performance.

\subsubsection{Training Efficiency of MIG-GT}

\begin{figure}[!tp]
\centering
\subfloat[
Baby\label{fig:speed_train_time_baby}
]{
\includegraphics[width=1.0in]{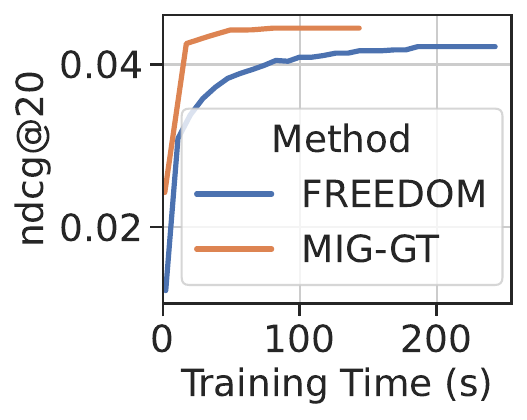}
}
\subfloat[
Sports\label{fig:speed_train_time_sports}
]{
\includegraphics[width=1.0in]{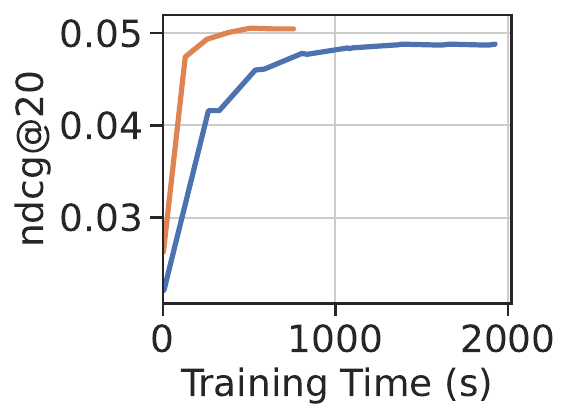}
}
\subfloat[
Clothing \label{fig:speed_train_time_clothing}
]{
\includegraphics[width=1.0in]{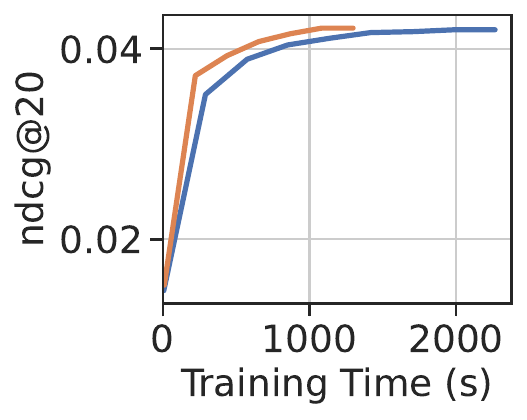}
}
\vspace{-1mm}
\caption{Test performance (ndcg@20) during training.}
\label{fig:speed_train_time} 
\vspace{-2mm}
\end{figure}

To demonstrate the training efficiency of our method, we visualize test performance (ndcg@20) against training time in seconds during training in Figure~\ref{fig:speed_train_time}.
The choice to report training time instead of epochs is deliberate. 
In recommendation tasks, the definition of an `epoch' can vary, being defined as a full iteration over either users (vertices) or user-item interactions (edges).
Thus, training time serves as a more consistent and comparable measure of efficiency across different methods.

When compared with FREEDOM -- known for its efficiency -- our method consistently achieves higher performance more rapidly across all tested graphs by avoiding a complex denoising mechanism over item-item relations.
On Baby and Sports, our method surpasses FREEDOM's final performance at early stages. 
On Clothing, while our final performance matches FREEDOM's, our method converges faster and reaches the optimal results earlier.

\subsubsection{Comparison with Contrastive Learning (CL)-Based Methods}
Another research direction focuses on improving GNNs via CL.
We build MIG-GT-CL, integrating our model with the typical CL loss, InfoNCE~\cite{oord2019infonce}, and compare it against CL-based methods, MMSSL~\cite{DBLP:conf/www/WeiHXZ23}, MGCN~\cite{DBLP:conf/mm/Yu0LB23}, and LGMRec~\cite{DBLP:conf/aaai/GuoL0WSR24}.
Results in Table~\ref{tab:exp_compare_cl_methods} show that MIG-GT already surpasses most baselines and outperforms all with the simple addition of the CL loss (MIG-GT-CL).

\begin{table}[!tp]
\centering
\scalebox{0.72}{
\begin{tabular}{l|cc|cc|cc}
\hline
 & \multicolumn{2}{c|}{Baby} & \multicolumn{2}{c|}{Sports} & \multicolumn{2}{c}{Clothing} \\
\hline
Method          &  R@20  & N@20   &  R@20 & N@20  &  R@20 & N@20  \\
\hline

MMSSL	    & 0.0971 & 0.0420 &	0.1013 & 0.0474 & 0.0797 & 0.0359 \\
MGCN	    & 0.0964 & 0.0427 &	0.1106 & 0.0496 & 0.0945 & \textbf{0.0428} \\
LGMRec	    & 0.1002 & 0.0440 &	0.1068 & 0.0480 & 0.0828 & 0.0371 \\ \hline
MIG-GT	    & 0.1021 & \textbf{0.0452} &	\textbf{0.1130} & \textbf{0.0511} & 0.0934 & 0.0422 \\
MIG-GT-CL	& \textbf{0.1022} & 0.0451 &	0.1120 & 0.0505 & \textbf{0.0946} & \textbf{0.0428} \\

\hline
\end{tabular}
}
\vspace{-2mm}
\caption{Comparison with CL-Based Methods.}
\label{tab:exp_compare_cl_methods}
\vspace{-2mm}
\end{table}

\section{Conclusions}

In this study, we explored GNNs for multimodal recommendation systems.
We observe that optimal receptive fields for GNNs can vary across different modalities. 
To capitalize on this, we introduced GNNs with Modality-Independent Receptive Fields, employing separate GNNs for each modality with independent receptive fields to enhance performance.
To address the challenge where the optimal receptive field size, which can be quite low, restricts GNNs' ability to capture global information, we proposed a Sampling-based Global Transformer (SGT). 
It utilizes uniform global sampling to more efficiently integrate global information within GNN frameworks.
Experiments show that the SGT improves performance even with a small number of sampled vertices, confirming sampling as an effective method for applying Global Transformers in multimodal recommendations.

\clearpage
\newpage

\section*{Acknowledgements}

This research is supported by the National Research Foundation, Singapore and Infocomm Media Development Authority under its Trust Tech Funding Initiative, the National Research Foundation, Singapore under its AI Singapore Programme (AISG Award No: AISG2-TC-2021-002), and the National Natural Science Foundation of China (Grant No: 62106262).

\bibliography{aaai25}

\end{document}